\documentclass[fleqn,12pt,twoside]{article}
\usepackage{espcrc1}
\usepackage{graphicx}
\usepackage[figuresright]{rotating}
\newcommand{\dd}{ {\tiny \textrm d}}

\newcommand{\AmS}{{\protect\the\textfont2
  A\kern-.1667em\lower.5ex\hbox{M}\kern-.125emS}}

\title{Quick View on Jet Tomography at RHIC Energies}

\author{G. G. Barnaf\"oldi\address[RMKI]{RMKI KFKI, P.O. Box 49, 
        Budapest 1525, Hungary;  \\ 
        ITOL, E\"otv\"os University, 
        P\'azm\'any P. 1/A, Budapest 1117, Hungary}  
        \thanks{Thanks the EPS Scholarship and OTKA U047552 for support
		of my participation, and for my co-authors, P. L\'evai, 
		G. Papp, G. Fai., for discussions, and advices. This 
		work was supported in part by Hungarian grants T034842, 
		T043455, T043514, U.S. DOE grant: DE-FG02-86ER40251, 
		and NSF grant: INT0000211. Supercomputer time provided 
		by BCPL in Norway and the EC -- Access to Research 
		Infrastructure action of the Improving Human Potential 
		Programme is gratefully acknowledged.}
}
\begin{document}
\maketitle

\begin{abstract}
A strong suppression of $\pi$ production in $AuAu$ collision 
was discovered at RHIC energies in the high-$p_T$ region. The 
lack of this suppression in $dAu$ collision proved, this 
effect can be connected to final state effects, namely 
medium induced jet energy loss caused by dense color matter.
Here we display our quantitative results on jet quenching 
in $AuAu$ collision using our pQCD-based parton model calculations.
In parallel we present results on $dAu$ collision.  
\end{abstract}


\section{Short introduction}
\label{intro}

Recent high-$p_T$ RHIC experimantal results on pion production 
in central $AuAu$ collision at mid-rapidity at $\sqrt{s}=130$ 
and $200$ AGeV have shown a strong suppression compared to binary 
scaled $pp$ data~\cite{phenix0203,star0203}. Analysing this effect in 
different impact parameter ranges, the suppression vanishes with 
increasing centrality. In $dAu$ the lack of the 
suppression~\cite{phenix0203} was found also. A quantitative 
analysis of the suppression pattern, 
the so called {\it "jet-tomography"}~\cite{tomog}, yields information 
about the physical aspects of the produced dense matter and the impact 
parameter dependence of the formation of deconfined matter in $AuAu$ 
collisions. Latest data on $dAu$ collision confirmed that initial 
state effects can not be responsible for suppression in $AuAu$ collision.
Thus medium induced non-Abelian jet-energy loss in the final state 
becomes a strong candidate to explain the missing pion yield.
Jet energy loss can be calculated in a perturbative quantum 
chromodynamics (pQCD) frame~\cite{glv}. Our aim is
to calculate $\pi$ production in nucleus-nucleus $AA$ collision in 
a pQCD improved parton model, taking into account a phenomenological 
intrinsic transverse momentum distribution (intrinsic-$k_T$) for the 
colliding partons, which is necessary to get better agreement 
between data and calculations in $pp$ collisions~\cite{Yi02,Bp02}. 
For $AA$ collision we included initial state effects into our calculations: 
nuclear multiple scattering, saturation in the number of semihard collisions, 
and shadowing effect inside nucleus. We used Glauber-like collision geometry.  

Here we investigate jet energy loss. As high-energy quark and gluon 
jets are travelling throught dense colored matter, they can lost their 
energy described by induced gluon radiation in thin non-Abelian matter. 
The GLV-description~\cite{glv} of energy loss can be included into the pQCD 
improved parton model as a strong final state  effect. 
Comparing theory to data we can extract the opacity 
values ($\bar{n}=L/\lambda$) of the produced colored matter at 
different centralities.

\section{Numerical results vs. experimental data }
\label{sec:2}

\begin{figure}[htb]
\vspace{-1.5truecm}
\begin{minipage}[t]{90mm}
\resizebox{90mm}{60mm}{\includegraphics{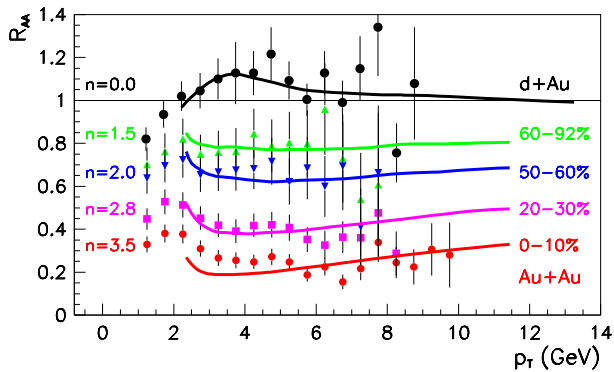}}
\vspace{-1.73truecm}
\caption{$R_{AA}$ for pions from $AuAu$ and $dAu$ collisions, data are  
taken from Refs.~\cite{phenix0203,star0203}, and calculation for 
$dAu$ from Ref.~\cite{dau}.}
\label{fig:1}
\vspace{-1.3truecm}
\end{minipage}
\end{figure}

Applying the above sketched model for reproducing $\pi$ yields
in $AuAu$, and $dAu$ collisions we can accomplish a tomographic analysis 
of the produced dense and colored matter in reactions. In Fig. \ref{fig:1} 
we display the nuclear modification factor: 
$R_{AA'}(p_T)=
\frac{E_{\pi} \dd \sigma^{AA'}_{\pi}/\dd^3 p_T}
{N_{bin}\,\,E_{\pi} \dd \sigma^{pp}_{\pi}/\dd^3 p_T},$
as a function of $p_T$ for pion production from central to peripheral 
$AuAu$ ({\sl lower curves}) and in $dAu$ ({\sl the upper curve}). 
In the most central $AuAu$ collision the measured suppression is reproduced
with opacity $L/\lambda =$ 3.5. As Fig. \ref{fig:1} displays, in $AuAu$ 
collisions the suppression of the high-energy pions goes away with decreasing 
centrality, which is followed by the descreasing values of the opacity.  
In $dAu$ collision we do not need to include jet quenching to reproduce pion 
production, this final state effect does not appear there, only 
multiscattering and shadowing.



\begin{thebibliography}{99}

\bibitem{phenix0203} 
     G. David {\it et al.} (PHENIX Coll.),
      Nucl. Phys. A698 (2002) 227; 
     D. d'Enterria {\it et al.} (PHENIX Coll.),
      Nucl. Phys. A715 (2003) 749;
     S.S. Adler {\it et al.} (PHENIX Coll.),
      Phys. Rev. Lett. 91 (2003) 072301, {\it ibid.} 072303;
     {\tt nucl-ex/0308006}.

\bibitem{star0203}
     J.C. Dunlop {\it et al.} (STAR Coll.),
      Nucl. Phys. A698 (2002) 515; 
     J.L. Klay     {\it et al.} (STAR   Coll.),
      Nucl. Phys. A715 (2003) 733;
     J. Adams {\it et al.} (STAR Coll.),
      Phys. Rev. Lett. 91  (2003) 172302, {\it ibid.} 072304;
    {\tt nucl-ex/0308023}.

\bibitem{tomog}
      E. Wang, X.N. Wang,
       Phys. Rev. Lett. 89 (2002) 162301;
      M. Gyulassy, P. L\'evai, I. Vitev, 
       Phys. Lett. B538 (2002) 282.

\bibitem{glv} 
     M. Gyulassy, P. L\'evai and I. Vitev,
     Phys. Rev. Lett.  85 (2000) 5535;
     Nucl. Phys.  B571 (2000) 197,
     {\it ibid.}  B594 (2001) 371.

\bibitem{Yi02}
     G. Papp {et al.} {\tt hep-ph/0212249};
     Y. Zhang {\it et al.},
     Phys. Rev. C65 (2002) 034903. 

\bibitem{Bp02}
    G.G. Barnaf\"oldi {\it et al.}, 
    APH NS Heavy Ion Phys. 18 (2003) 79, {\tt nucl-th/0206006}. 

\bibitem{dau}
    P. L\'evai. {\it et al.}, {\tt nucl-th/0306019}.

\end{thebibliography}
\end{document}